\documentclass[12pt]{article}
\usepackage{epsfig,amssymb}
\usepackage{latexsym}

\newcommand{\bq}{\begin{equation}}
\newcommand{\eq}{\end{equation}} \newcommand{\bqa}{\begin{eqnarray}}
\newcommand{\eqa}{\end{eqnarray}} \newcommand{\ben}{\begin{enumerate}}
\newcommand{\een}{\end{enumerate}}
\newcommand{\bc}{\begin{center}}
\newcommand{\ec}{\end{center}} \newcommand{\bqb}{\begin{eqnarray*}}
\newcommand{\eqb}{\end{eqnarray*}}

\def\pr#1#2#3{ Phys. Rev. ${\bf{#1}}$:#2 (#3)}
\def\prl#1#2#3{ Phys. Rev. Lett. ${\bf{#1}}$:#2 (#3)}

\def\np#1#2#3{ Nucl. Phys. ${\bf{#1}}$:#2 (#3)}

\def\ie{{\it i.e. }}
\def\eg{{\it e.g. }}

\global\nulldelimiterspace = 0pt

\begin{document}
\pagenumbering{arabic}
\thispagestyle{empty}
\def\thefootnote{\fnsymbol{footnote}}
\setcounter{footnote}{1}

\begin{flushright}
April   2006\\
PTA/06-7\\
hep-ph/0604041\\
Corrected version \\

 \end{flushright}
\vspace{2cm}
%---------------------titre ---------------------------------------
\begin{center}
{\large\bf Addendum to "Helicity conservation in gauge boson scattering
at high energy"}.
\vspace{1.cm}  \\
%-----------------------------------------------------------------
{\large G.J. Gounaris$^a$ and F.M. Renard$^b$}\\
\vspace{0.2cm}
$^a$Department of Theoretical Physics, Aristotle
University of Thessaloniki,\\
Gr-54124, Thessaloniki, Greece.\\
\vspace{0.2cm}
$^b$Laboratoire de Physique
Th\'{e}orique et Astroparticules, UMR 5207,\\
Universit\'{e} Montpellier II,
 F-34095 Montpellier Cedex 5.
\vspace*{1.cm}

{\bf Abstract}
\end{center}
\vspace*{-0.4cm}
In a previous paper  we have established that for any
two-body process involving   even numbers
of transverse  vector bosons and  gauginos, the dominant asymptotic  amplitudes
obey helicity conservation (HC); \ie the sum of the helicities of the two incoming
particles equals to the sum of the helicities of the two outgoing ones.
 This  HC has been proved   to all orders  in MSSM,
 provided  $(s,~ |t|,~|u|)$ are much
larger than all  masses in the model; but only
to leading 1-loop logarithmic order in SM.
In the present addendum, the validity of HC is extended
to all two-body processes.
Renormalizability is crucial for the HC validity,
while all known anomalous couplings violate it.\\


\def\thefootnote{\arabic{footnote}}
\setcounter{footnote}{0}
\clearpage

Asymptotic helicity conservation (HC)  states
that  the dominant   helicity amplitudes $F(a_{\lambda_1}b_{\lambda_2}
 \to c_{\lambda_3} d_{\lambda_4})$ for any two-body process
\bq
a_{\lambda_1}+b_{\lambda_2}\to c_{\lambda_3}+d_{\lambda_4} ~, ~ \label{2body-process}
\eq
involving an even number of transverse gauge bosons,
should satisfy
\bq
\lambda_1+\lambda_2=\lambda_3 +\lambda_4 ~~, \label{hc-rule}
\eq
with   $\lambda_j$ denoting the ordinary helicities,
and  $(s, |t|,|u|)$  being  much larger than all  masses in   either the SM
or the MSSM model \cite{HCprl}. Particularly for MSSM, the validity of
the HC theorem  demanded also that the number of  participating
gauginos is again   even\footnote{Gaugino-higgsino or sfermion mixings
always disappear at high energies, where HC applies.}.

 In case  both initial (final) particles have spin $1/2$,
while the final (initial) ones are bosons, the relation
\bq
\lambda_1+\lambda_2=\lambda_3 +\lambda_4=0 ~~, \label{hc-even-rule}
\eq
should   also  be obeyed  by the asymptotically dominant amplitudes  \cite{HCprl}.
It is obvious that (\ref{hc-even-rule}) would had been impossible,
if the number of transverse gauge bosons  were not  even.

This HC theorem  has been proved in \cite{HCprl}
to one-loop leading logarithmic
order in SM, but  to all orders  in MSSM.
For   MSSM in particular, the proof implied that all amplitudes
violating (\ref{hc-rule}),
should vanish at energies much larger
than all supersymmetric particle masses \cite{HCprl}.

 The very interesting property  of  HC is that  it  strongly
 reduces the number of independent amplitudes at high energy.
As an example we quote the processes
$e^-+ e^+  \to W^- + W^+$ in SM or MSSM, where the  validity of
(\ref{hc-rule}, \ref{hc-even-rule}) forces the dominant  amplitudes
at high energies to have  opposite helicities, for both, the
$W^\mp$ and the $e^\mp$ pairs. \\

In the present addendum, the HC theorem is extended  to
processes involving   an odd number  of
transverse gauge bosons,  or an odd    number  of
 gauginos. Combining this with the results of \cite{HCprl}, we
obtain therefore that  HC should be valid for  all    two-body
processes.   According to this,
 the  dominant amplitudes  at asymptotic energies, in either SM or MSSM,
  should   always satisfy (\ref{hc-rule});
  while  the additional  validity of  (\ref{hc-even-rule})
   only holds  for   processes involving   an even number
 of transverse gauge bosons.  As in \cite{HCprl}, this extended
 HC theorem continues to be  valid  to one-loop leading logarithmic
order in SM, while  in  MSSM  it  holds to all orders at high energies. \\

To make our point, we first look at  various examples
of   processes  not covered by  \cite{HCprl}, for which one-loop
complete or leading logarithmic calculations exist in the literature \cite{BRV}.

The first concerns  $g+ b \to W^- +t$
which has been studied in either SM or MSSM,
to one-loop leading logarithmic order  \cite{BMRV-odd}.
For transverse $W^-$, it of course obeys the HC rule
(\ref{hc-rule}) \cite{HCprl}.
But for longitudinal $W^-$ also, it has  been found in \cite{BMRV-odd},
that the asymptotically dominant
amplitudes   obey (\ref{hc-rule}),  to the above  order.
This is also true for $g+ b \to H^- +t$ occurring  in  MSSM \cite{BMRV-odd}.

As  a second  example we consider the processes
\bq
 q_{L(R)}+g \to \tilde \chi_i^0 + \tilde q_{L(R)}~~,~~
 u_{L(R)}+g \to \tilde \chi_i^+ +\tilde d_{L(R)}~~, \label{gq-chisq}
\eq
 involving  one  transverse gauge boson. Both these processes have been found to
 respect\footnote{The   processes in (\ref{gq-chisq}) are of even order
 in the fermion-fermion-Higgs Yukawa couplings, and thus obey
 the U(1) symmetry discussed below.}  the HC rule (\ref{hc-rule}),
 to the one-loop leading logarithmic
order \cite{single}. A similar property has also been observed
by an explicit calculation for
$ q_{L(R)}+g \to \tilde g + \tilde q_{L(R)}$.

Finally, as a third   example we quote
\bq
\tilde \chi_i^0+ \tilde \chi_j^0 \to \gamma +Z ~~,  \label{nn-gamZ}
\eq
which describes the annihilation of any pair of neutralinos to $\gamma Z$.
The full  one-loop calculation of this process  has been
recently completed in MSSM, for any energy and scattering angle \cite{Diakonidis}.
In performing this calculation, it has been observed that
the dominant asymptotic amplitudes for a transverse $Z$ respect both
(\ref{hc-rule}, \ref{hc-even-rule}), as expected from  \cite{HCprl};
while  only (\ref{hc-rule}) is obeyed  for a longitudinal $Z$.\\

We next turn to the all-order   proof of (\ref{hc-rule}) in MSSM,
 following   the same method as in \cite{HCprl}, and  working in the exact
supersymmetric limit, with  a vanishing  Higgs-mixing parameter $\mu$.
All particles are massless in this limit, and  a  new global U(1) symmetry appears,
whose charge we call "formal helicity"   denoted  by $\bar \lambda_i$
\cite{HCprl}. By definition, $\bar \lambda_i$ vanishes  for the gauge fields,
equals to the normal helicity  for the fermion fields,
 to -1 for the L-sfermion and the Higgs fields, and to +1 for the R-sfermion
fields \cite{HCprl}.
All terms in this   model are invariant
under the above   U(1), except the fermion-fermion-scalar interactions
induced by the Yukawa terms in the superpotential, which  violate it
 by 2 units.

Since for vanishing masses,   crossing of
a particle from an  incoming to an  outgoing state always
changes its helicity,
the only independent two body processes we need to consider
consist of the SUSY related pairs
\bqa
       \tilde V +\tilde f_{L(R)} \to \tilde H + \tilde f'_{R(L)} &, &
     V + f_{L(R)} \to  (H,~G) +  f'_{R(L)} ~~,  \label{gaugino-sf} \\
\tilde V +f_{L(R)} \to \tilde H + f'_{R(L)} &, &
     V +\tilde f_{L(R)} \to ( H,~G) + \tilde f'_{R(L)} ~~,  \label{gaugino-f}
\eqa
involving an odd number of gauginos or an odd number of transverse gauge bosons,
and  the process
\bqa
 && V + f_{L(R)} \to \tilde V' +  \tilde f'_{L(R)}
  ~~,  \label{gauge-gaugino-f}
\eqa
containing simultaneously one  transverse gauge and one gaugino.
In (\ref{gaugino-sf},
\ref{gaugino-f}, \ref{gauge-gaugino-f}), $f$ denotes fermions, $\tilde f$ sfermions,
$\tilde H$ higgsinos, $(H,~G)$ Higgs or Goldstone bosons, $V$ transverse gauge bosons,
and $\tilde V$ gauginos. The indicated  chiralities of
the  fermions and sfermions,  are the only ones allowed for
 the above amplitudes to be non vanishing in the considered model.
 Since we restrict to two-body processes,
 the aforementioned  requirement that the number
of  participating transverse gauge or gauginos  is odd,
just means that this number
can never be larger than
one\footnote{This is because the same treatment as the one presented
in the  paragraphs below,
also leads to the conclusion that the two-body processes involving
three (transverse) gauge and one scalar boson,
or one  gauge and three scalars,
always vanish asymptotically, in agreement with HC.}.

Turning now to the proof of (\ref{hc-rule}),
we first discuss the left-side process in (\ref{gaugino-sf}).
The diagrams contributing to  it  involve at most pairs
of hermitian conjugates of the above Yukawa terms,
multiplied by a single  such  term,   inducing   a U(1)
violation by two units of   "formal helicity", \ie
\bq
\Big |\bar \lambda_{\tilde V}+\bar \lambda_{\tilde f}-
\bar \lambda_{\tilde H}-\bar \lambda_{\tilde f'} \Big |=2~~. \label{formal-1}
\eq
Analyzing  the possible discrete values of
the "formal"   and the corresponding     ordinary helicities,
we  find that the  asymptotically dominant
amplitudes for the left process in (\ref{gaugino-sf}) obey
$|\bar \lambda_{\tilde f}-\bar \lambda_{\tilde f'}|=2$,
while the ordinary helicities  satisfy
\bq
 \lambda_{\tilde V} =  \lambda_{\tilde H}
=\pm \frac{1}{2} ~~~ {\rm for } ~~~ \tilde f_{L(R)} ~ ~~ \label{HC-gsb-Hst}
\eq
for the two $\tilde f$-chiralities respectively.
As a result, equation   (\ref{hc-rule}) is respected by the dominant
amplitudes of the left process in (\ref{gaugino-sf}).

Performing now a  SUSY transformation to this process,
under which
the gaugino $\tilde V$  is transformed to a transverse gauge boson $V$
carrying a helicity of the same
sign \cite{HCprl},  and $(\tilde f_{L(R)}, \tilde f'_{R(L)})$ are
 transformed to $(f_{L(R)}, f'_{R(L)})$,
we end up with the right process in (\ref{gaugino-sf})
\bq
 V + f_{L(R)} \to  (H,~G) +  f'_{R(L)} ~~, \label{gf-Hfprime}
\eq
where the ordinary   helicities of the asymptotically dominant
 amplitudes   satisfy $\lambda_V+\lambda_f = \lambda_{f'}$, in agreement
with\footnote{Note that the correspondence in the
 right part of (\ref{HC-gsb-Hst}) guarantees that the $V$-helicity
 in  (\ref{gf-Hfprime})
  must be of  opposite sign to that of the  f-helicity.
 Therefore  the $V$ and $f'$  helicities carry the same
 sign in this process.} (\ref{hc-rule}).

Using then   the equivalence theorem for the Goldstone case,
  we end up with the validity of (\ref{hc-rule})
for the  longitudinal $W$ or $Z$-production process \cite{equiv}
\bq
 V + f_{L(R)} \to  (W_{\rm long },~ Z_{\rm long })+   f'_{R(L)}~~, \label{gb-Wt}
\eq
depending on the $G$ charge.
The validity of  (\ref{hc-rule}) for  transverse $W$ or $Z$,
 has already been established in  \cite{HCprl}.

As an illustration of the (\ref{gaugino-sf}) case, we may quote
\bq
\tilde g + \tilde b_{L(R)} \to \tilde H^- + \tilde t_{R(L)} ~~,
~~ g + b_{L(R)} \to  (G^-,H^-) +  t_{R(L)}~~, \label{gsb-Hst}
\eq
where in the left process a  gluino and a b-squark annihilate producing  a higgsino
and a stop.

An identical treatment  may  be also done  for the processes
in (\ref{gaugino-f}).

An analogous  study may also be made for the process  (\ref{gauge-gaugino-f})
whose  diagrams respect U(1), since
they always involve  pairs of hermitian conjugates of the above Yukawa terms.
The sum of the "formal helicities" in the initial and final states are therefore
equal, for  the asymptotically dominant amplitudes;   which
in turn again  leads to the validity of (\ref{hc-rule})
for the ordinary helicities.
For reaching this conclusion,  it is important
to remember that the gauge and gaugino fields connected by
SUSY transformations, always carry helicities of the same sign \cite{HCprl}.
Thus, HC is valid also
for amplitudes  simultaneously involving one  transverse gauge  boson
and one  gaugino.
As  an illustration  we quote the rather academic case
\bq
W^- + u_{L(R)} \to \tilde W^0 + \tilde d_{L(R)} ~~~. \label{gauge-gaugino}
\eq

The HC asymptotic rule (\ref{hc-rule}) has therefore
been established for all processes
 in (\ref{gaugino-sf}, \ref{gaugino-f}, \ref{gauge-gaugino-f}).
Combining this with the results of \cite{HCprl}, we conclude  that
HC holds asymptotically for any  two-body process.
For this to be true  though,  we must always stay away from
the infrared and collinear singularities of
 the two-body amplitudes; since  these
  singularities are intimately related to the
   multibody processes, for which HC is known  not to apply \cite{Dixon}.\\

It is  worth emphasizing that the renormalizability of the underlying theory
is  crucial  for the validity of HC. All
 anomalous (non-renormalizable) couplings  we have looked at,
 violate  HC  \cite{Kazimierz}.  An experimental
program to test  the high energy HC property,   would therefore provide  an
 important simple check of the nature
of the basic interactions.

\vspace{0.5cm}
\noindent{\large\bf{Acknowledgement}}\\
\noindent
Work supported by the European Union
under contracts HPRN-CT-2000-00149.
G.J.G. gratefully acknowledges also the support by the European Union
 contract MRTN-CT-2004-503369.

\newpage

\end{document}